\documentclass[modern]{aastex63}

\usepackage{amsmath}
\usepackage{amssymb}

\usepackage{soul} 
\usepackage{wasysym}

\shorttitle{Interior of Bennu}
\shortauthors{Tricarico et al.}

\begin{document}

\title{Internal rubble properties of asteroid (101955) Bennu}

\correspondingauthor{P. Tricarico}
\email{tricaric@gmail.com}

\author[0000-0001-6540-512X]{P. Tricarico}
\affiliation{Planetary Science Institute, Tucson, AZ, USA}
\affiliation{Independent Researcher, Beaucouz\'e, France}

\author[0000-0003-0558-3842]{D. J. Scheeres}
\author[0000-0002-3427-1920]{A. S. French}
\author[0000-0002-1847-4795]{J. W. McMahon}
\author{D. N. Brack}
\affiliation{Smead Department of Aerospace Engineering Sciences, University of Colorado, Boulder, CO, USA}

\author[0000-0003-3632-6793]{J. M. Leonard}
\author[0000-0001-9040-310X]{P. Antreasian}
\affiliation{KinetX Aerospace Inc., Simi Valley, CA, USA}

\author[0000-0003-3240-6497]{S. R. Chesley}
\author[0000-0003-0774-884X]{D. Farnocchia}
\author[0000-0003-1444-3454]{Y. Takahashi}
\affiliation{Jet Propulsion Laboratory, California Institute of Technology, Pasadena, CA, USA}

\author[0000-0003-3456-427X]{E. M. Mazarico}
\author{D. Rowlands}
\author{D. Highsmith}
\author[0000-0002-7253-5931]{K. Getzandanner}
\author[0000-0002-4699-0519]{M. Moreau}
\affiliation{NASA Goddard Spaceflight Center, Greenbelt, MD, USA}

\author{C. L. Johnson}
\affiliation{Planetary Science Institute, Tucson, AZ, USA}
\affiliation{Department of Earth, Ocean and Atmospheric Sciences, University of British Columbia, Vancouver, BC, Canada}

\author[0000-0002-5286-8528]{L. Philpott}
\affiliation{Department of Earth, Ocean and Atmospheric Sciences, University of British Columbia, Vancouver, BC, Canada}

\author[0000-0001-5890-9821]{E. B. Bierhaus}
\affiliation{Lockheed Martin Space, Littleton, CO, USA}

\author[0000-0002-0906-1761]{K. J. Walsh}
\affiliation{Southwest Research Institute, Boulder, CO, USA}

\author[0000-0002-3578-7750]{O. S. Barnouin}
\affiliation{Johns Hopkins University Applied Physics Laboratory, Laurel, MD, USA}

\author[0000-0001-6755-8736]{E. E. Palmer}
\author[0000-0002-2830-1708]{J. R. Weirich}
\author[0000-0002-2293-7879]{R. W. Gaskell}
\affiliation{Planetary Science Institute, Tucson, AZ, USA}

\author[0000-0002-3733-2530]{M. G. Daly}
\author[0000-0002-2163-7276]{J. A. Seabrook}
\affiliation{Centre for Research in Earth and Space Science, York University, Toronto, ON, Canada}

\author[0000-0001-8316-0680]{M. C. Nolan}
\author[0000-0002-2597-5950]{D. S. Lauretta}
\affiliation{Lunar and Planetary Laboratory, University of Arizona, Tucson, AZ, USA}

\begin{abstract}
Exploration of asteroid (101955) Bennu by the OSIRIS-REx mission
has provided an in-depth look at this rubble-pile near-Earth asteroid.
In particular, the measured gravity field and the detailed shape model of Bennu
indicate significant heterogeneities in its interior structure,
compatible with a lower density at its center.
Here we combine gravity inversion methods with a statistical rubble-pile
model to determine the density and size-frequency distribution (SFD) index
of the rubble that constitutes Bennu.
The best-fitting models indicate that the SFD of the interior is consistent with that observed on the surface,
with a cumulative SFD index of approximately $-2.9$.
The rubble bulk density is approximately $1.35$~g/cm$^3$, corresponding to a $12$\% macro-porosity.
We find the largest rubble particle to be approximately $145$~m, whereas the largest void is approximately $10$~m.
\end{abstract}

\section{Introduction} \label{sec:intro}

The small ($\sim 500$~m diameter) near-Earth asteroid (101955) Bennu was explored in detail by the Origins, Spectral Interpretation, Resource Identification, and Security--Regolith Explorer (OSIRIS-REx) mission \citep{2019Natur.568...55L}.
The surface presents a wide range of particle sizes and craters \citep{2019NatAs...3..341D,2019NatGe..12..242W}
The top-like shape of Bennu indicates a rubble-pile structure \citep{2019NatGe..12..247B} --- that is, loosely bound fragments of rock that gravitationally reaccumulated following the destruction of a larger parent body ---
providing initial evidence for some degree of internal mass density heterogeneity \citep{ScheeresNatAstro2019}.

The discovery of particle ejection events on Bennu’s surface \citep{2019Sci...366.3544L}
and the reconstruction of the particles’ trajectories \citep{2020JGRE..12506363C}
has led to a very accurate measurement of the gravity field of Bennu,
allowing us to study in detail its interior properties in terms of mass density distribution.
\citet{Scheereseabc3350} presented the results of a first analysis,
employing analytical as well as numerical techniques,
showing that the deviation of Bennu’s interior from a homogeneous mass density distribution is consistent with
lower density at the center of the asteroid and along its equatorial bulge.

Here we combine gravity inversion methods with a statistical rubble-pile
model to determine the density and size-frequency distribution (SFD) index
of the rubble that constitutes Bennu.

\section{Methods} \label{sec:methods}

As we return to study the interior density distribution of Bennu,
we need to remember that in general the inverse gravity problem is intrinsically under-determined
and admits an infinite family of solutions \citep{2013GeoJI.195..260T}.
In this context, assumptions about the properties of the solution are typically used
to focus on certain properties of the density distribution.
The observed gravity of Bennu is close to that generated by a homogeneous density distribution (inhomogeneities detected through degree 3) \citep{2020JGRE..12506363C}.
Therefore, the natural question that we aim to answer here is how close to homogeneous can the density
distribution be and still generate the observed gravity field.
Studying density solutions as close as possible to homogeneous
represents the simplest possible assumption that we can make about the solutions.
This approach has several advantages.
First, because the nominal homogeneous solution is unique,
we restrict the infinite family of solutions to a corner of the solution space,
where we would not normally find any solutions unless the data
allow them there.
Second, any features and deviations that we observe in the solutions
descend directly from the shape and gravity data and are not the results of specific additional assumptions made.

The approach followed here has two main components.
The first consists of a gravity inversion technique where the interior is modeled using a smooth density gradient.
This technique is applied to the shape and gravity data of Bennu to search for interior solutions as close as possible to homogeneous,
allowing us to determine the deviation from homogeneous of the interior of Bennu at different scales.
The second component is statistical modeling of a rubble pile to determine how the rubble particle SFD
and density determine the deviation from homogeneous at different scales.
By combining these two techniques, we can determine the rubble properties that
are compatible with the observed gravity and shape of Bennu.

\subsection{Global Gravity Inversion} \label{sec:GGI}

The gravitational potential of an arbitrary body 
is typically described by an expansion in spherical harmonics
\citep{1966tsga.book.....K,1995geph.conf....1Y}:
\begin{equation}
 U = \frac{GM}{r} \sum_{l=0}^{\infty} \sum_{m=0}^{l} \left( \frac{r_0}{r} \right)^l P_{lm}(\cos(\theta)) \left( C_{lm} \cos (m \phi) + S_{lm} \sin (m \phi) \right) \label{eq:U}
\end{equation}
where $G$ is the gravitational constant,
$M$ is the mass of the body,
$\{ r, \theta, \phi \}$ are the body-fixed barycentric spherical coordinates,
$r_0$ is the radius of a sphere completely enclosing the body, and
$P_{lm}$ is the associate Legendre function.
The Stokes coefficients $C_{lm}$ and $S_{lm}$ characterize the gravitational potential outside the $r_0$ sphere
and depend on the shape and mass density distribution of the body;
they can be determined by integrating over the volume $V$ of the body \citep{1995geph.conf....1Y}:
\begin{align}
C_{lm} &= \frac{(2-\delta_{m,0})}{M} \frac{(l-m)!}{(l+m)!} \int_V \rho(r,\theta,\phi) \left( \frac{r}{r_0} \right)^{l} P_{lm}(\cos(\theta)) \cos (m\phi) \ \text{d}V \label{eq:Clm} \\
S_{lm} &= \frac{(2-\delta_{m,0})}{M} \frac{(l-m)!}{(l+m)!} \int_V \rho(r,\theta,\phi) \left( \frac{r}{r_0} \right)^{l} P_{lm}(\cos(\theta)) \sin (m\phi) \ \text{d}V \label{eq:Slm}
\end{align}
where the volume element in spherical coordinates is $\text{d}V = r^2 \sin(\theta) \ \text{d}r \ \text{d}\theta \ \text{d}\phi$.
The $J_l=-C_{l0}$ terms are referred to as zonal coefficients,
while the generic $C_{lm}$ and $S_{lm}$ terms with $l>0$ are referred to as the tesseral coefficients.
Typically in the inverse problem the shape is known,
and we aim to determine the range of mass distributions that generate the observed gravitational field.

The global gravity inversion (GGI) approach
\citep{2013GeoJI.195..260T,2018NatGe..11..819T}
is used here to determine the range of interior density models
that are compatible with the observed shape and gravity field.
The density is modeled using smooth density functions,
with a half-wavelength resolution of approximately $2 r_0/l$,
and the solutions found depend not only on the input data,
but also on the assumptions we make.
As the evidence is that the interior of Bennu is close to homogeneous \citep{2020JGRE..12506363C},
we focus on exploring solutions that are as close as possible to homogeneous,
while still reproducing the observed gravity field,
by targeting solutions that minimize the cost function $f_\text{cost}=\sigma_\rho/\mu_\rho$
of the standard deviation of the density over the average density.
The covariance of the gravity coefficients
is included in the GGI computations by repeatedly sampling $C_{lm}$ and $S_{lm}$
in agreement with their covariance matrix \citep{2020JGRE..12506363C},
and then solving each time for the GGI density distribution closest to homogeneous.

When targeting solutions close to homogeneous,
most of the density variations are concentrated near the surface of a body,
especially for higher degree $l$.
This is due to the $(r/r_0)^l$ term in Eq.~\eqref{eq:Clm} and Eq.~\eqref{eq:Slm},
that progressively suppresses the contribution of regions near the center of a body as $l$ increases.
So in order to estimate the effective deviation from homogeneous at a given degree,
we need to sample regions close to the surface and progressively avoid regions in the deep interior.
The volume average of the $(r/r_0)^l$ term is:
\begin{equation}
\frac{\int (r/r_0)^l \ \text{d}V}{\int \text{d}V} = \frac{\int_{0}^{r_0} (r/r_0)^l \ r^2 \ \text{d}r}{\int_{0}^{r_0} r^2 \ \text{d}r} = \frac{3}{l+3} \label{eq:avg_r}
\end{equation}
so we only include a fraction $3/(l+3)$ of the total body volume closest to the surface
at each degree $l$ when computing the statistics $\mu_\rho$ and $\sigma_\rho$.
The relevant volume of the body corresponds to points with radius larger than
effective sensitivity radius $r^* = r_0^* [ l/(l+3) ]^{1/3}$,
where $r_0^*$ is the radius of a sphere with the same volume as the body.
This also allows us to define an average effective sensitivity depth as
$h^* = r_0^* - r^*$.

\subsection{Rubble Size-Frequency Distribution and Sampling} \label{sec:rubble}

As we model the interior of a rubble pile, we have several free parameters.
The rubble has a nominal bulk density $\rho$ and a cumulative SFD index $\alpha$ (negative).
The density $\rho$ is constrained to be larger than the body bulk density $\rho_b$,
with the ratio $V_r=(\rho_B/\rho)V_b$ setting the fraction of the body volume $V_b$ occupied by rubble,
and the remaining volume $V_v=V_b-V_r$ consisting of voids between rubble particles.
The rubble population is divided into $N$ bins, each bin $j \in [ 1, N ]$ corresponding to rubble with size:
\begin{equation}
    s_j = s_1 \left( \frac{s_N}{s_1} \right)^{(j-1)/(N-1)}
\end{equation}
where $s_1$ is the smallest size and $s_N$ is the largest size.
The cumulative SFD of the population is:
\begin{equation}
    P_j = \left( \frac{s_N}{s_1} \right)^{\alpha(j-N)/(N-1)}
\end{equation}
so that $P_N=1$, and the number of rubble particles with size $s_j$ is $P_j-P_{j+1}$.
Typically in simulations we fix $s_1$ to be sufficiently small compared to the scales we are testing,
and then determine $s_N$ iteratively in order to conserve the total volume occupied by rubble.
The rubble population volume is:
\begin{equation}
    V_P = \sum_{j=1}^{N} (P_j-P_{j+1}) s_j^3
\end{equation}
where $s_j^3$ is the volume of each rubble particle,
and the iterative correction to the largest rubble particle is $s_N \leftarrow s_N \sqrt[3]{V_r/V_P}$,
which typically converges very quickly.
The construction for the void population is equivalent to that of the rubble.

Once the rubble and void populations are built, we can use the multivariate hypergeometric distribution (MHD)
to sample a total of $n$ elements from the rubble and void populations without replacement,
where $n$ can go up to the total number of rubble particles and voids.
An efficient way to perform this sampling is to use the Dirichlet distribution to approximate the MHD,
see \citet{10.1016/S0167-9473(00)00007-4}.
To each random sample $n$ we can then associate a total volume, mass, and density,
and the distribution of these densities, with mean $\mu$ and standard deviation $\sigma$,
can then be compared to what we observe from the GGI modeling.
The variance is small when large volumes are considered, and then increases for smaller volumes.
To estimate the nominal volume corresponding to a given degree $l$,
we have the contribution from the $(r/r_0)^l$ term producing a factor $3/(l+3)$,
and if we consider that most of the gravitational signal is typically concentrated in the zonal coefficients,
especially for even degrees \citep{2020JGRE..12506363C},
then the $P_{l0}$ term with its $l$ zeros further subdivides the volume into $(l+1)$ parts.
The two contributions produce an effective volume:
\begin{equation}
    V_l = \frac{3}{(l+1)(l+3)} V_0  \label{eq:volume_degree_rubble}
\end{equation}
where $V_0$ is the body volume.
Typically when sampling the MHD, we obtain a nearly continuous range of possible volumes
and include only the volumes that are within $\pm 5$\% of $V_l$ in the statistical computations.

\section{Results} \label{sec:results}

\begin{figure*}
\centering
\includegraphics*[width=0.40\textwidth]{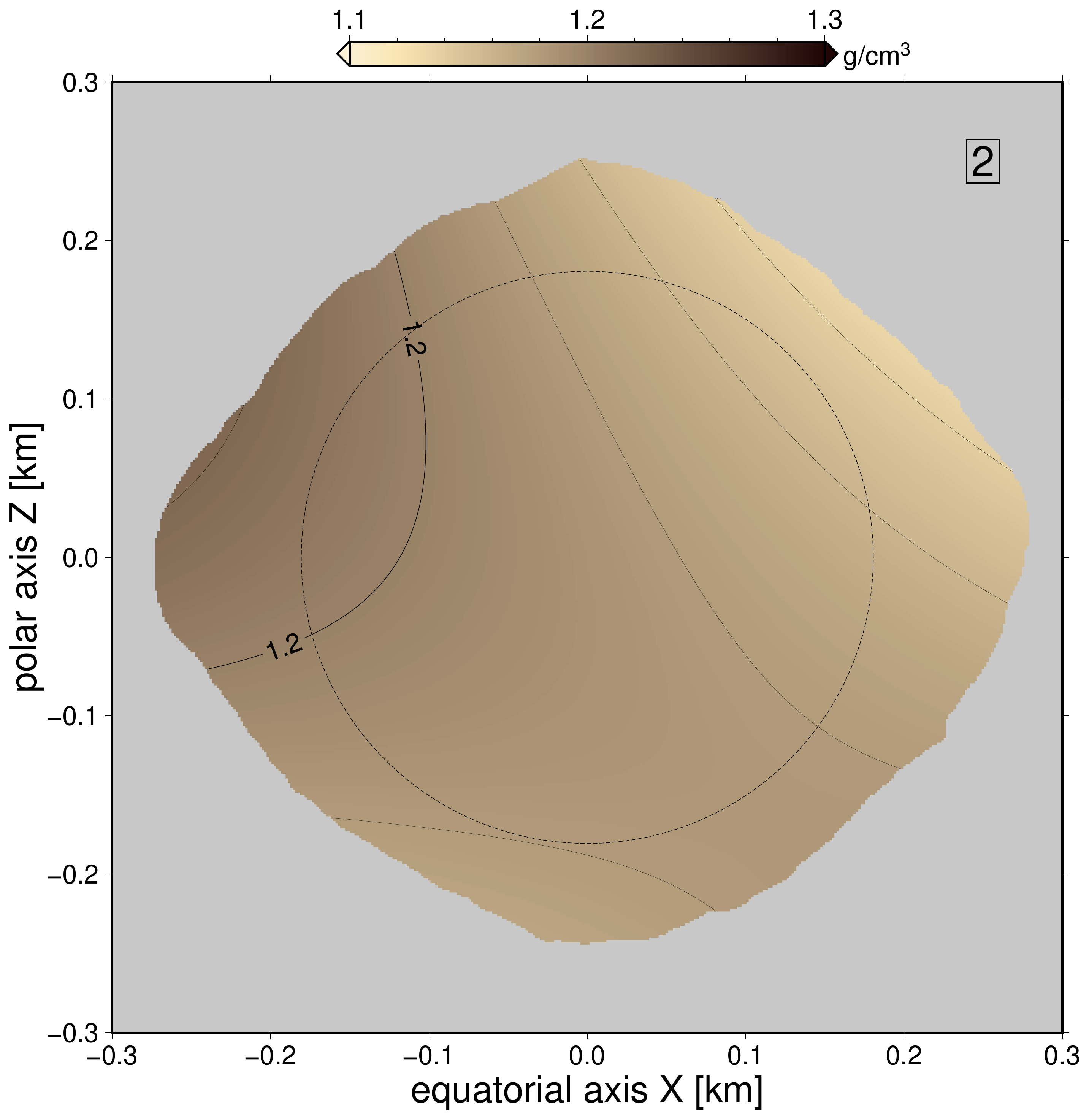}
\includegraphics*[width=0.40\textwidth]{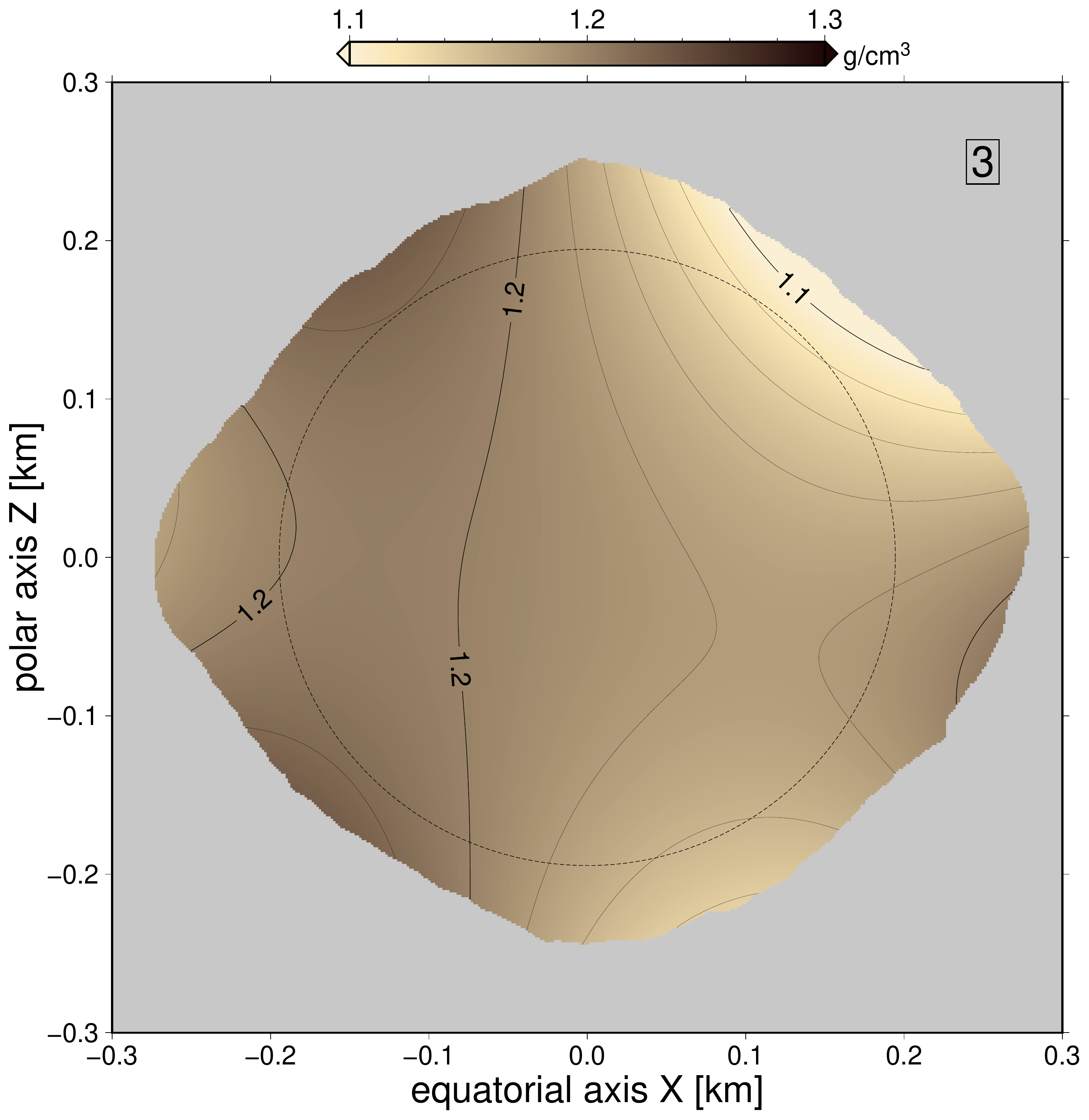}
\includegraphics*[width=0.40\textwidth]{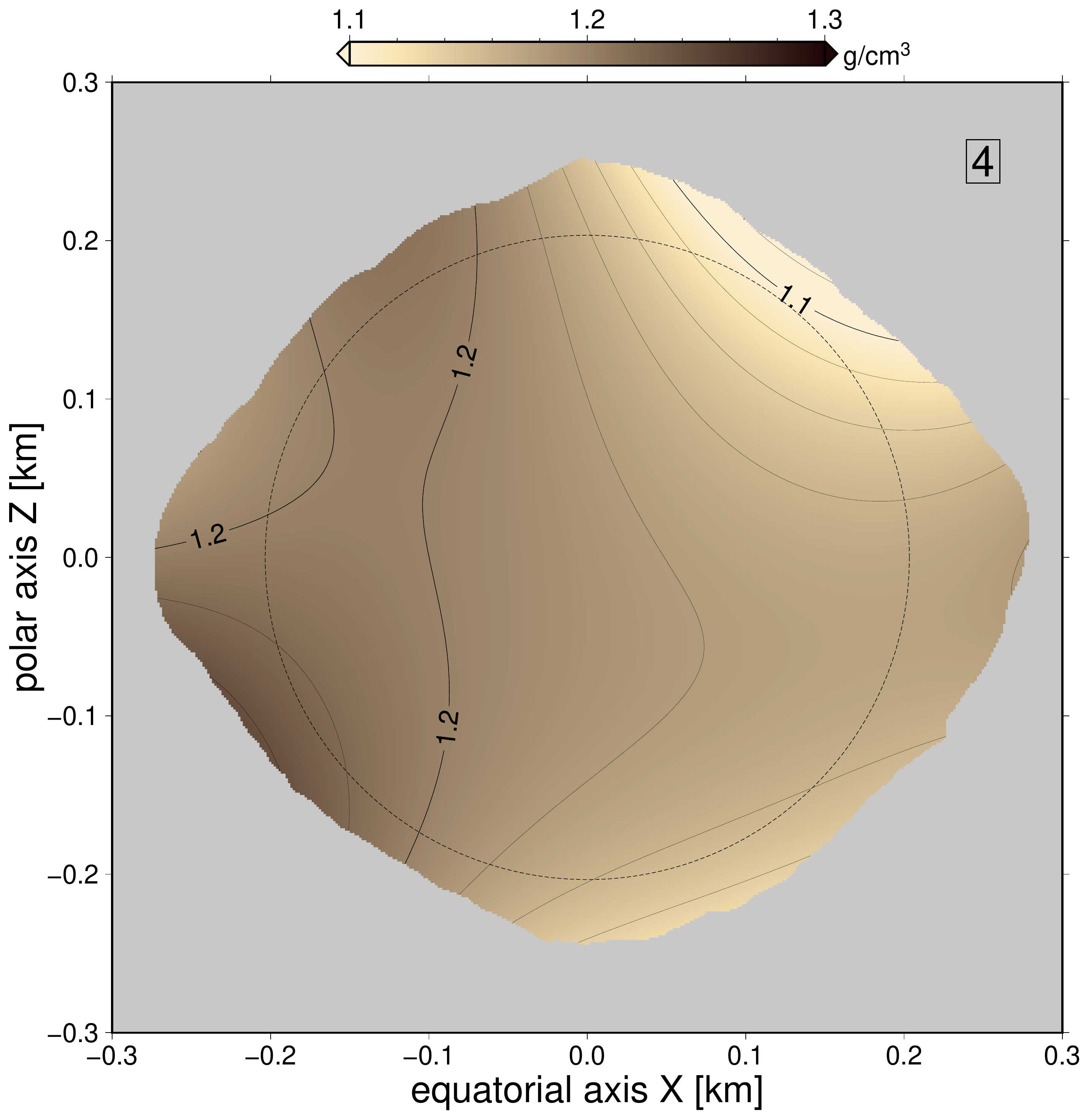}
\includegraphics*[width=0.40\textwidth]{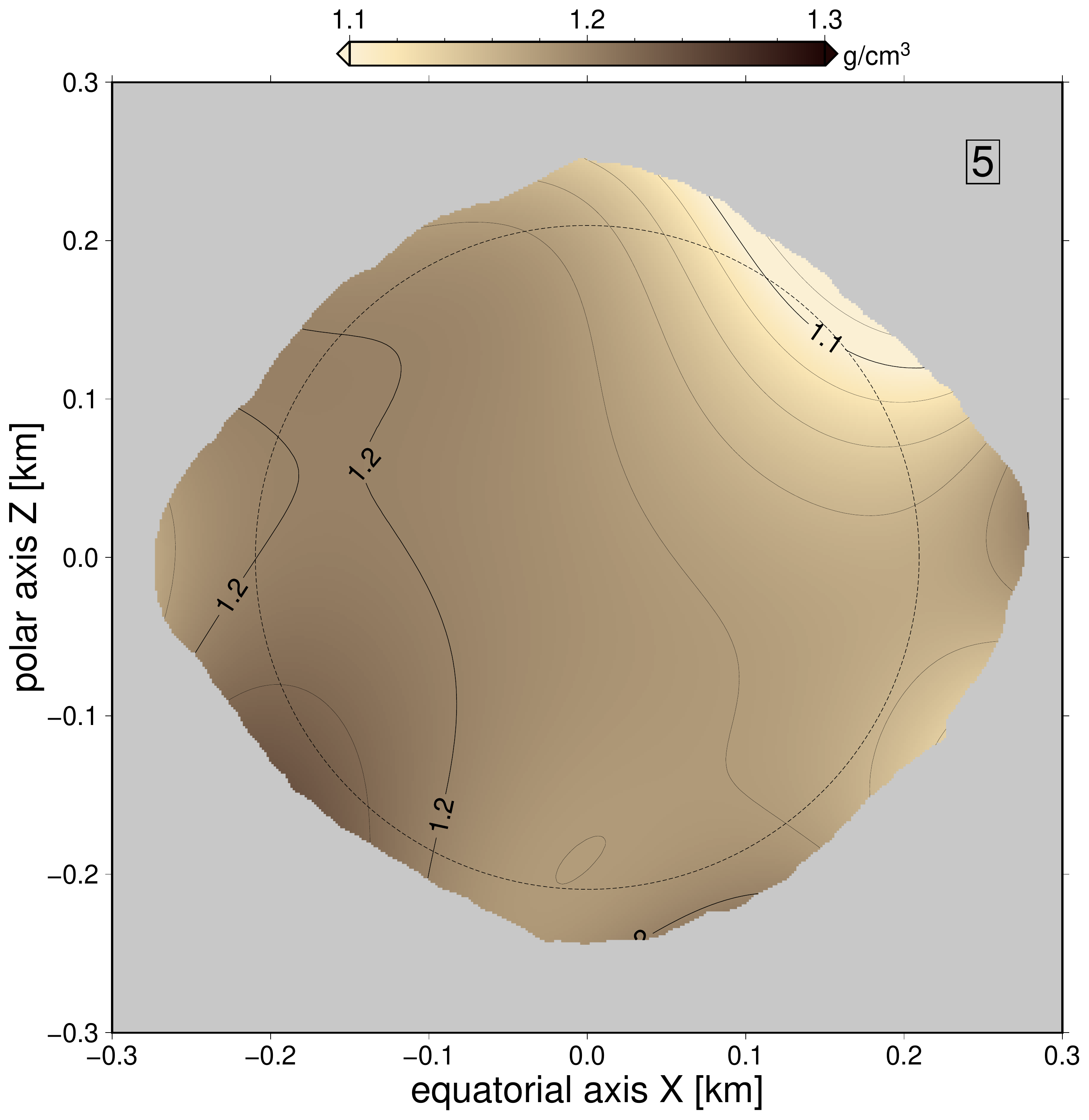}
\includegraphics*[width=0.40\textwidth]{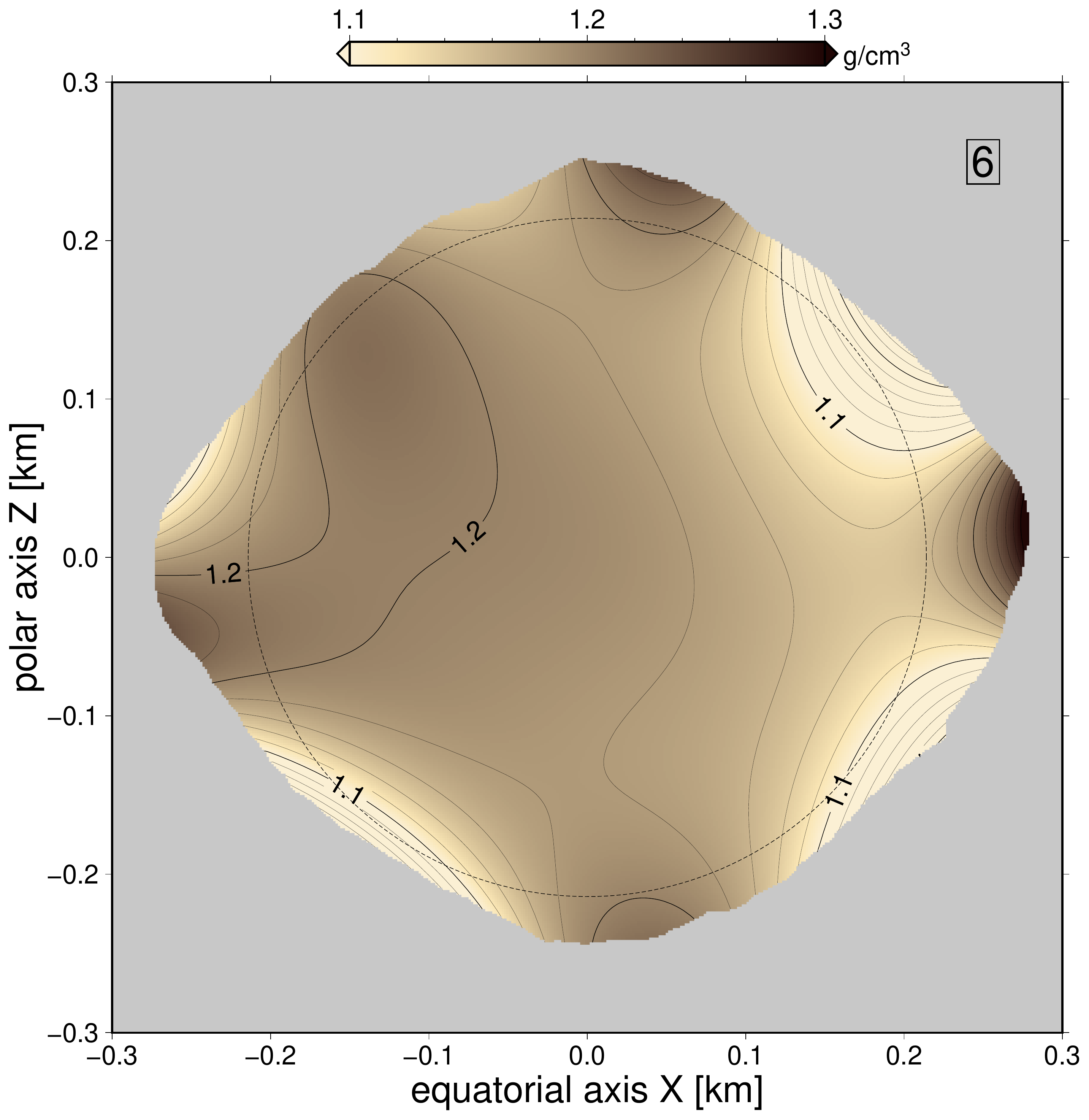}
\includegraphics*[width=0.40\textwidth]{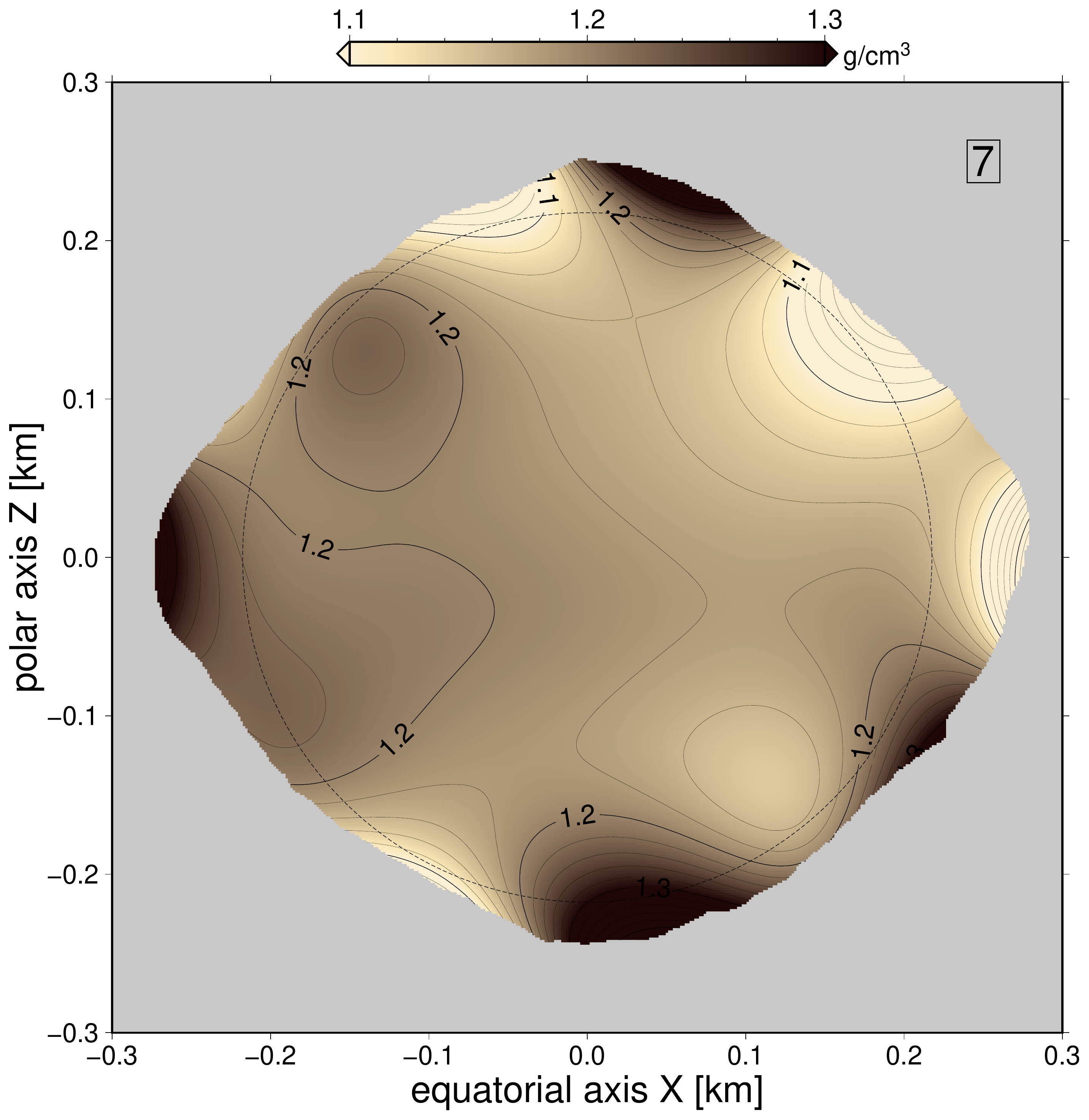}
\caption{
{\bf Interior density solutions for Bennu -- Polar Section.}
Solutions obtained by applying the GGI technique to the Bennu shape and gravity data.
Each diagram represents the interior density on the X--Z plane,
and is obtained including gravity data up to the degree indicated near the top right corner.
These solutions are representative of how close to homogeneous the interior of Bennu can be at each degree,
with the dashed line indicating the effective sensitivity radius.
}
\label{fig:slice_XZ}
\end{figure*}

\begin{figure*}
\centering
\includegraphics*[width=0.40\textwidth]{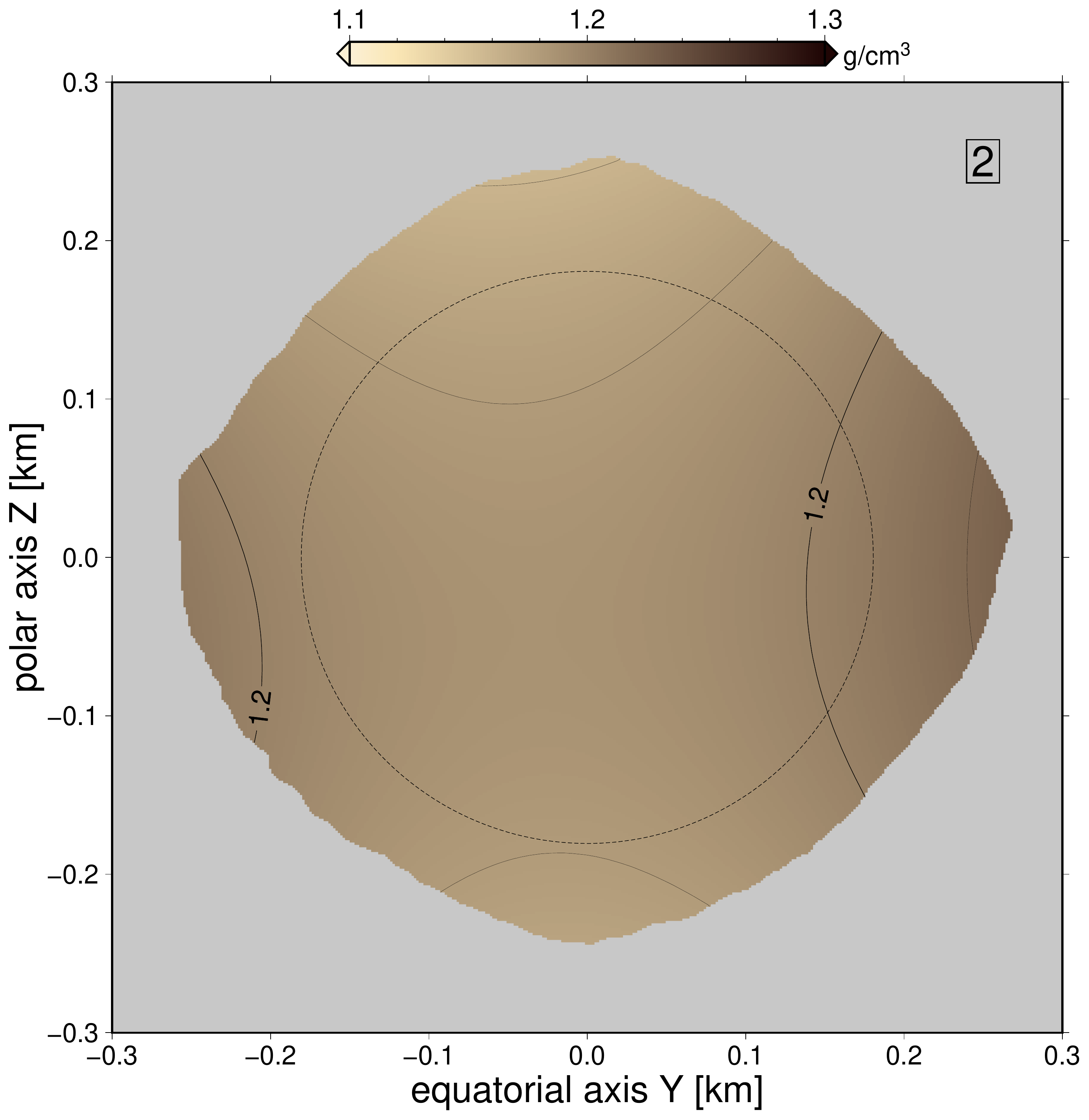}
\includegraphics*[width=0.40\textwidth]{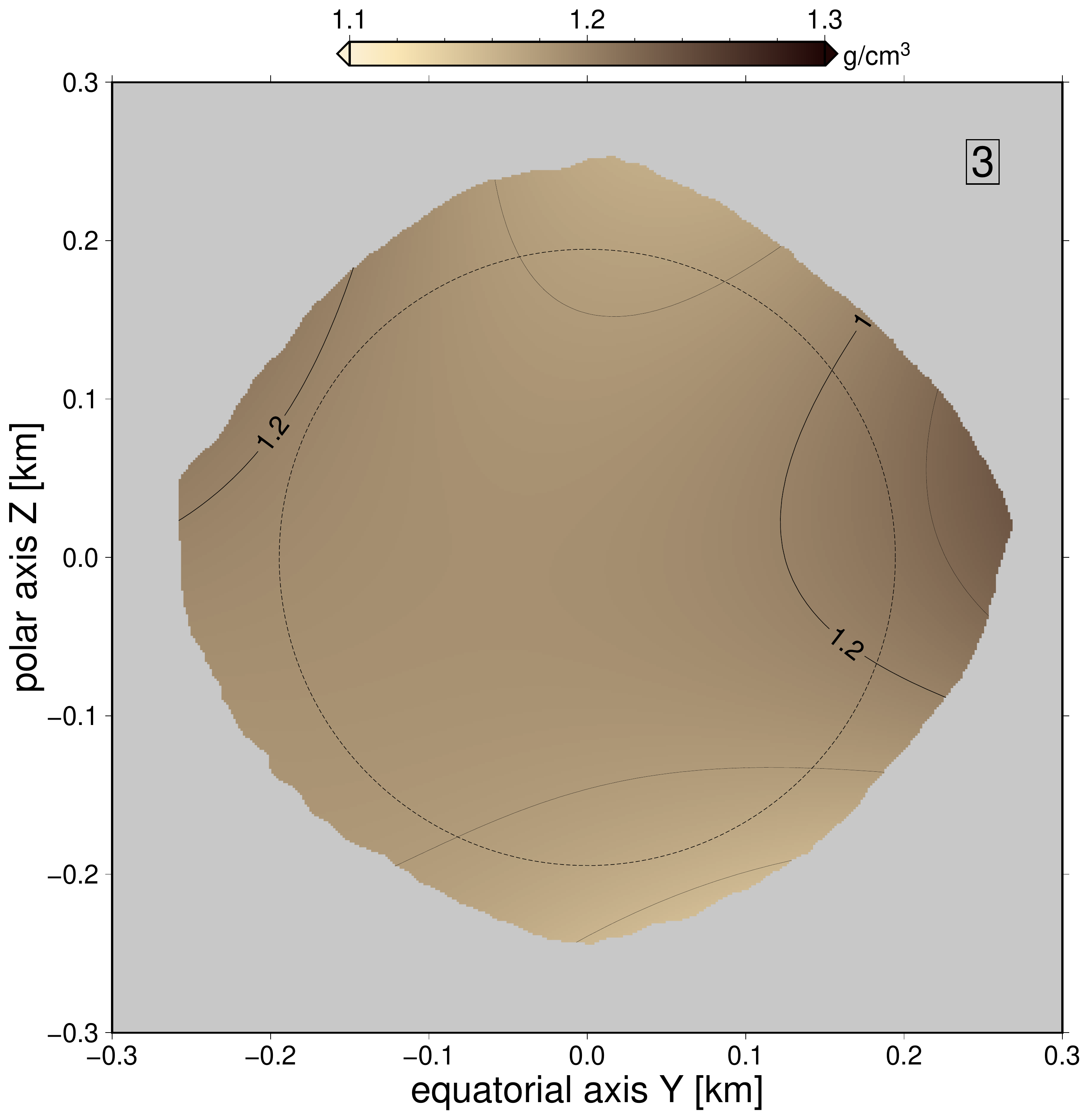}
\includegraphics*[width=0.40\textwidth]{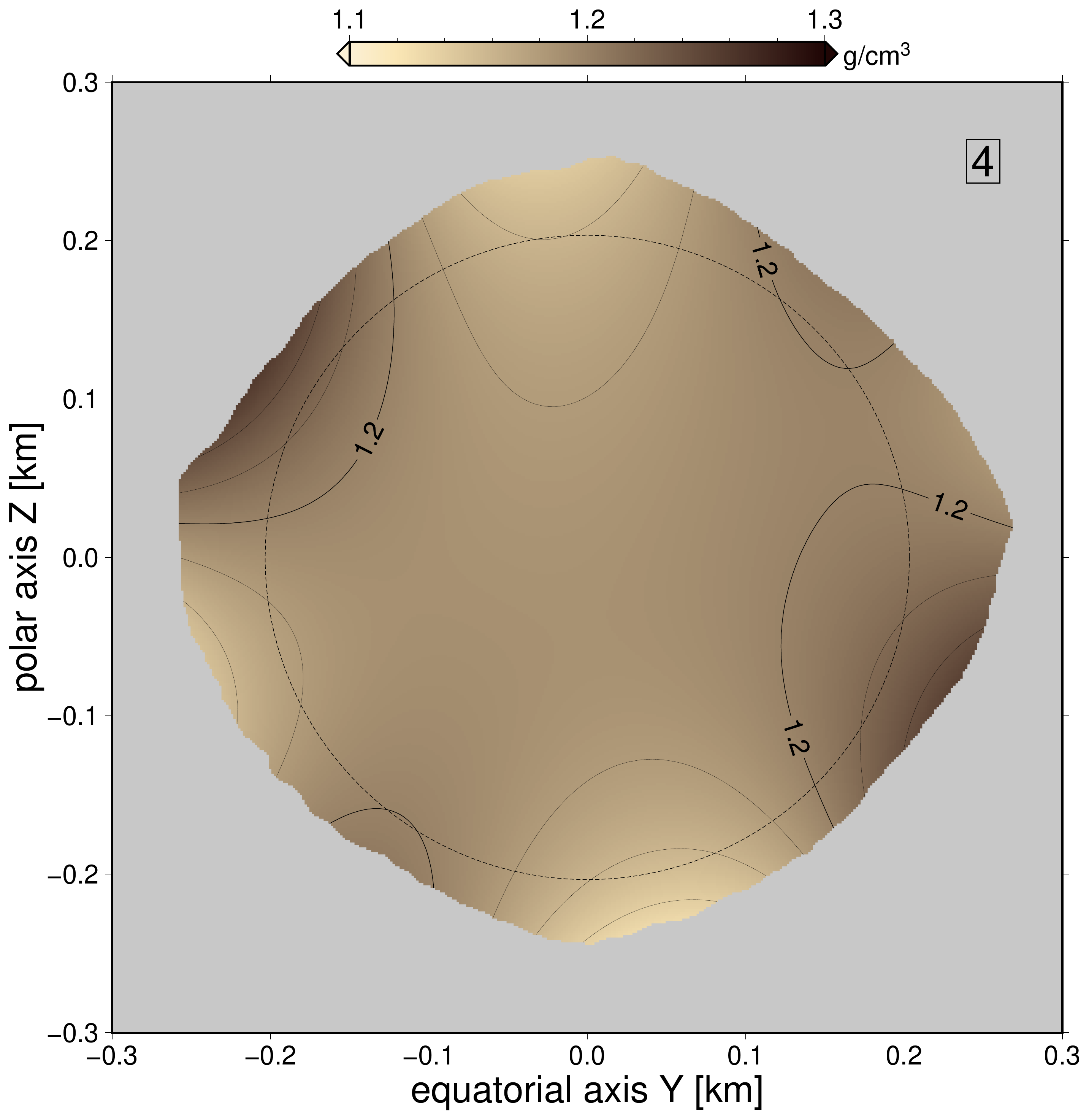}
\includegraphics*[width=0.40\textwidth]{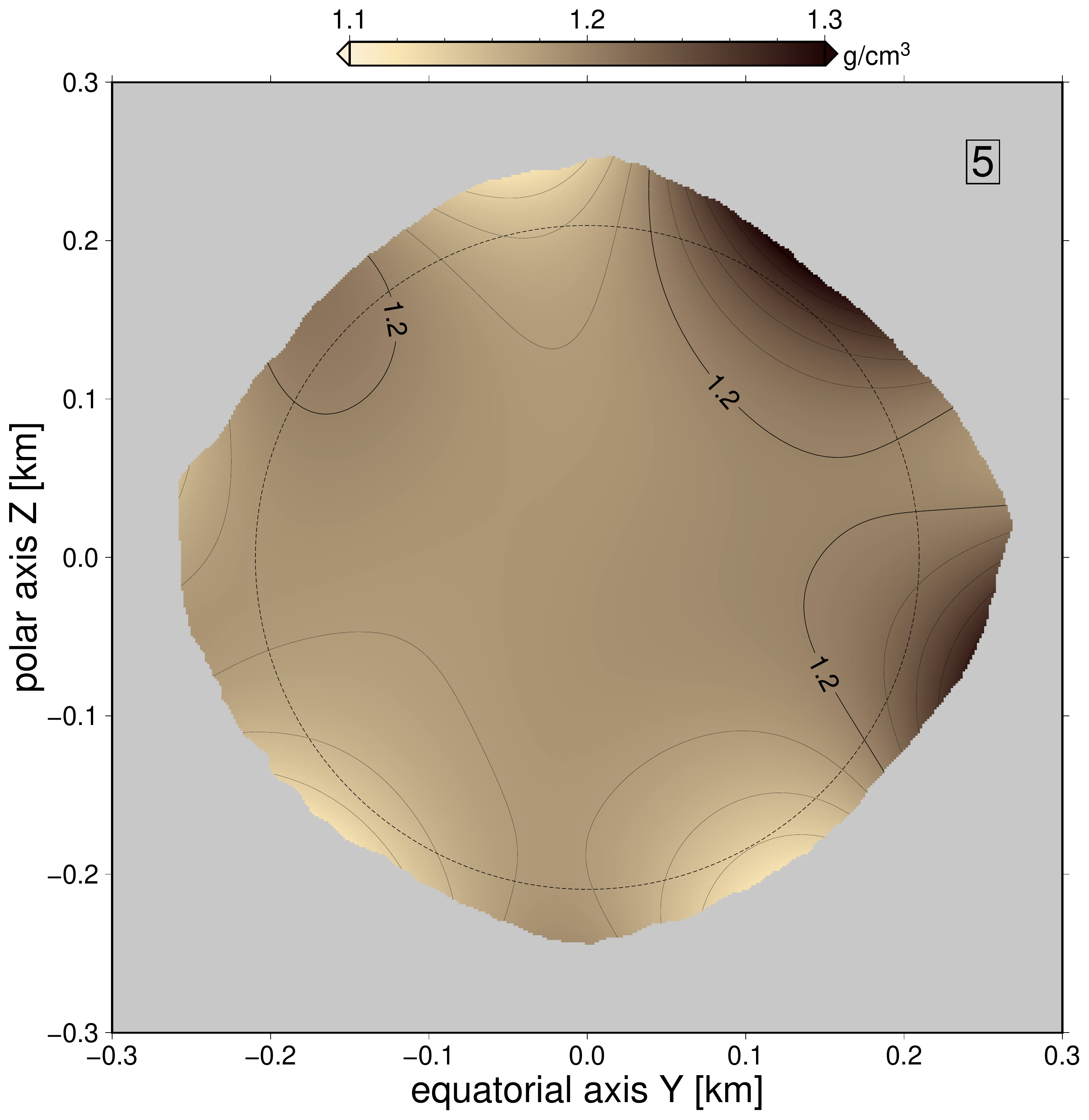}
\includegraphics*[width=0.40\textwidth]{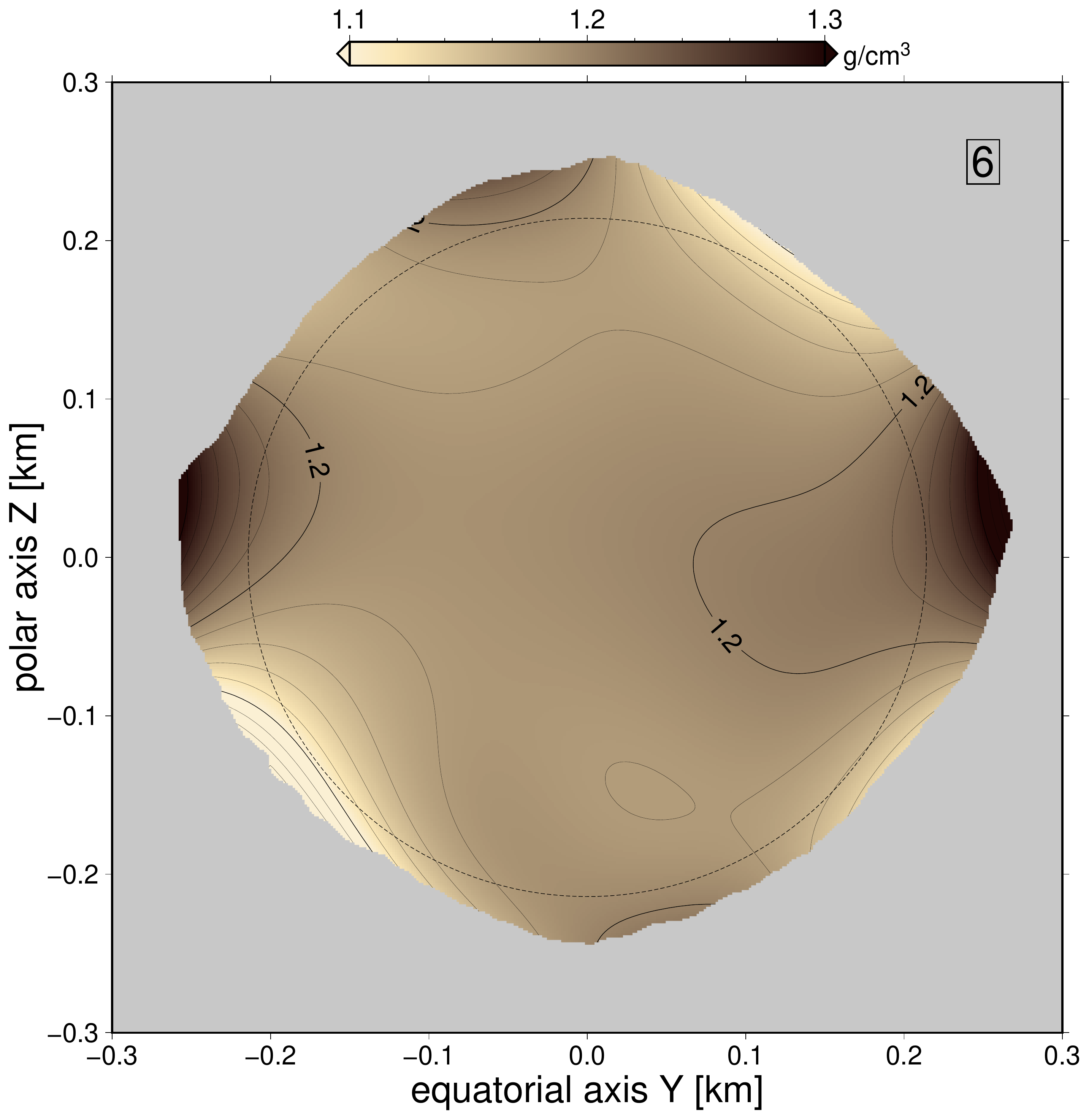}
\includegraphics*[width=0.40\textwidth]{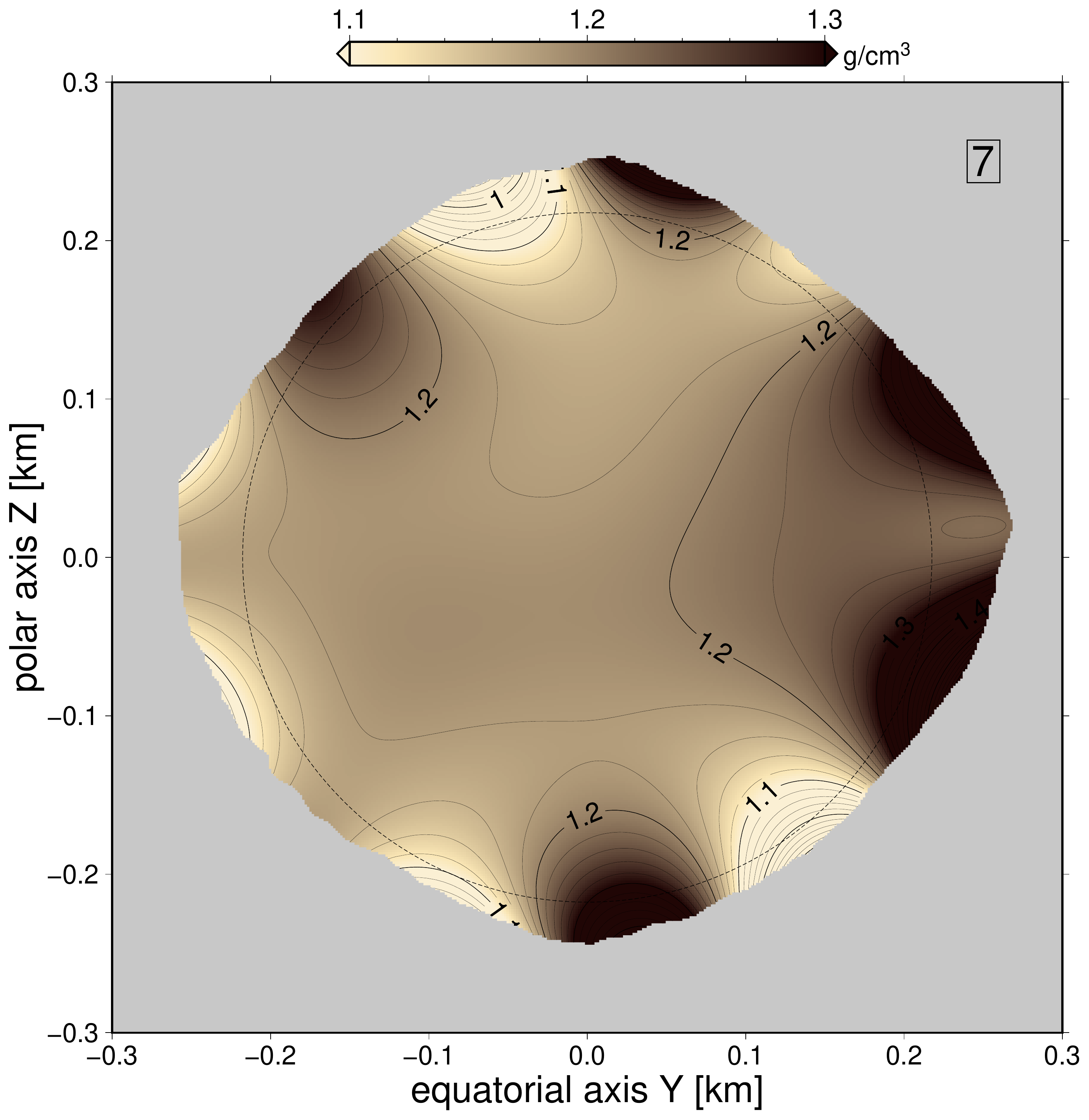}
\caption{
{\bf Interior density solutions for Bennu -- Alternate Polar Section.}
Same solutions as Figure~\ref{fig:slice_XZ}, but for the Y--Z plane.
}
\label{fig:slice_YZ}
\end{figure*}

\begin{figure*}
\centering
\includegraphics*[width=0.40\textwidth]{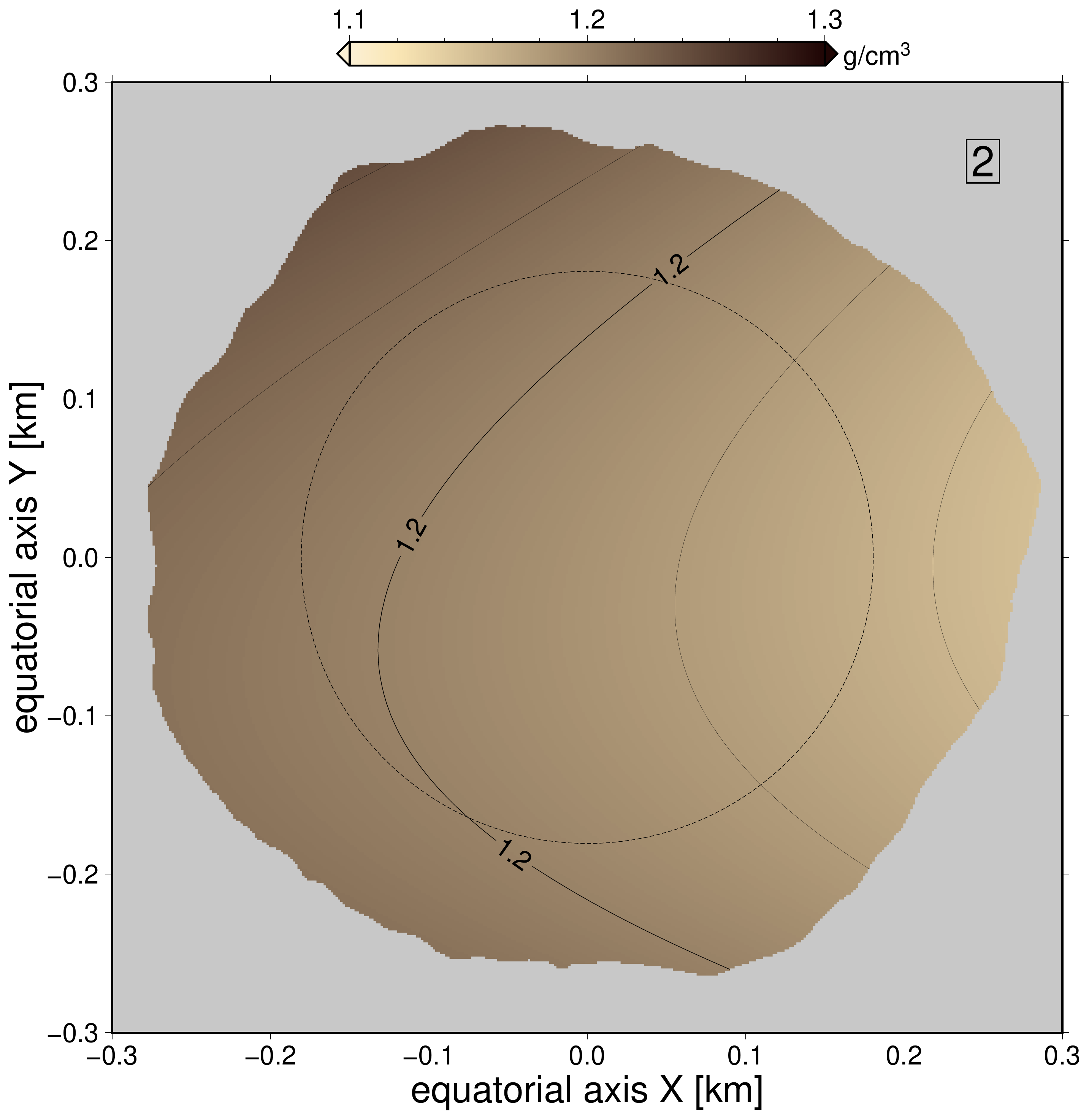}
\includegraphics*[width=0.40\textwidth]{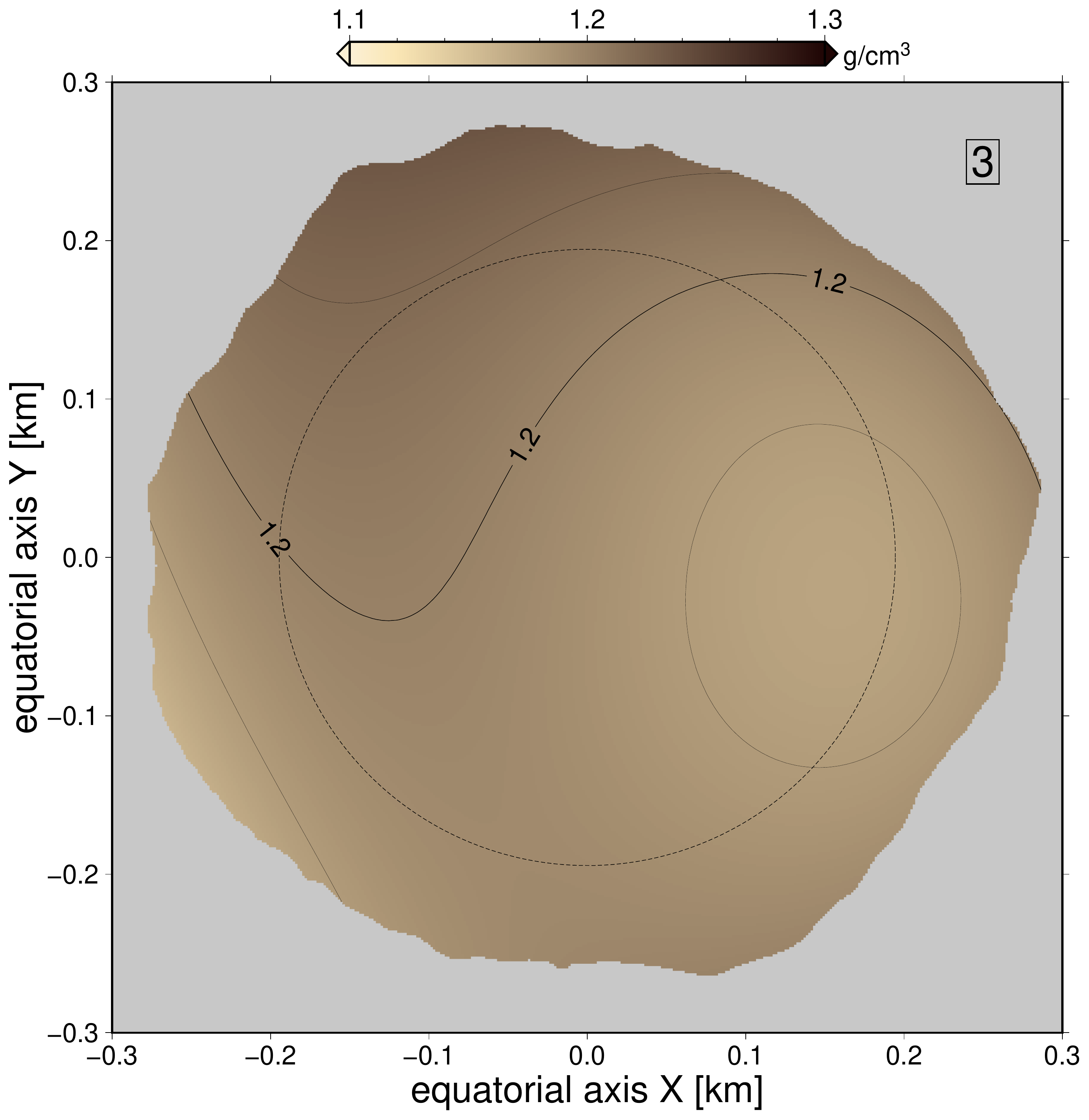}
\includegraphics*[width=0.40\textwidth]{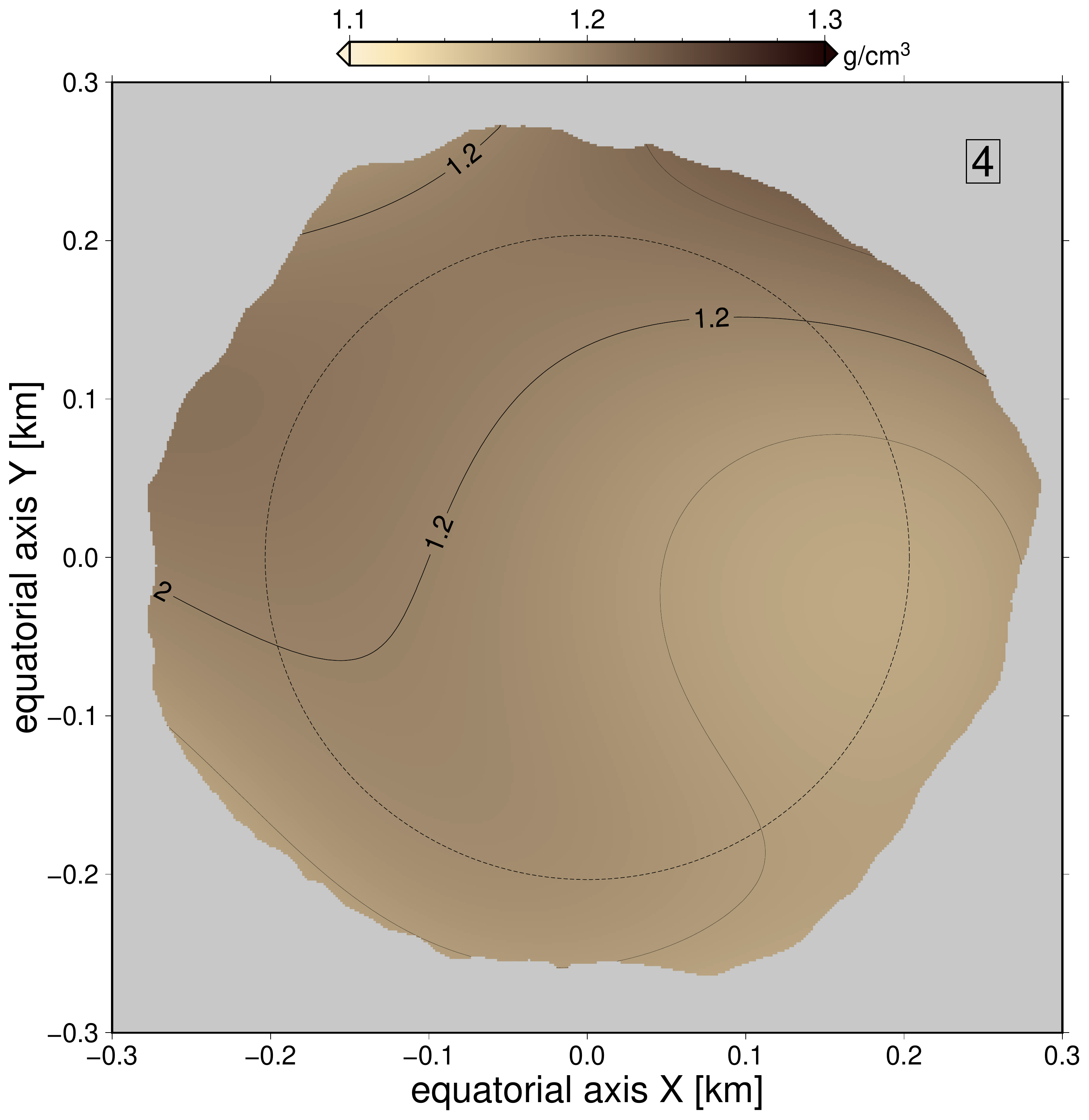}
\includegraphics*[width=0.40\textwidth]{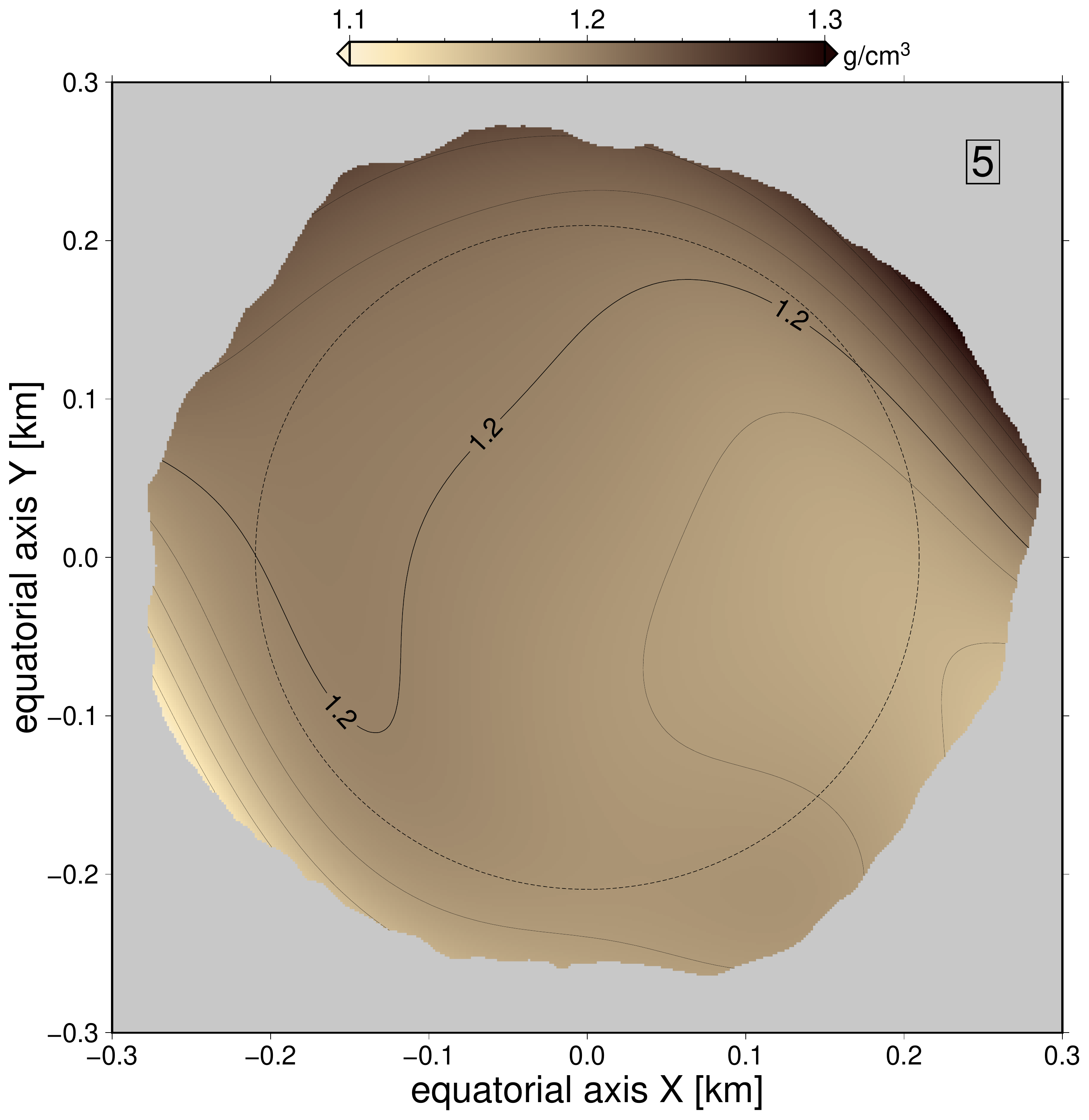}
\includegraphics*[width=0.40\textwidth]{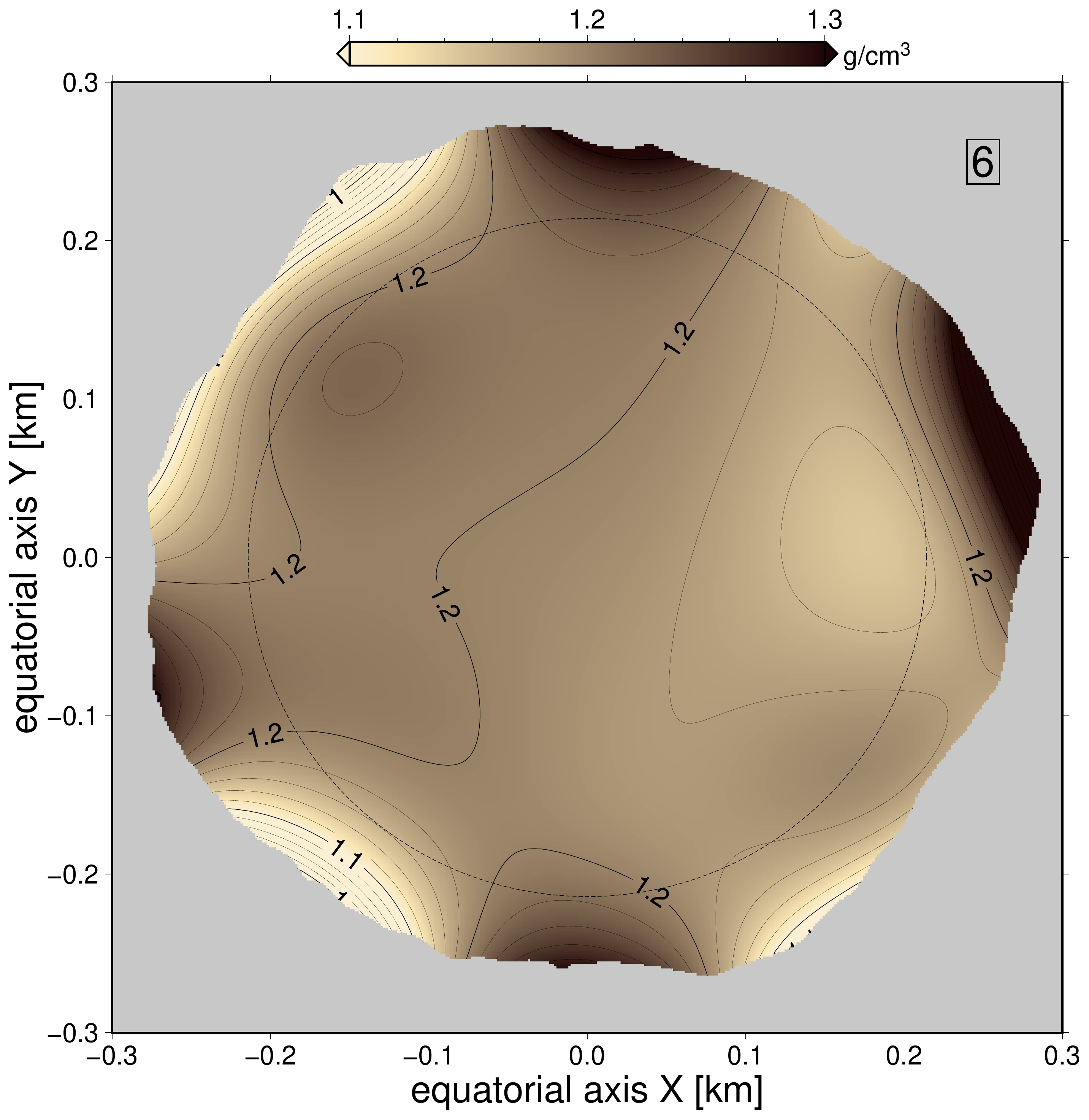}
\includegraphics*[width=0.40\textwidth]{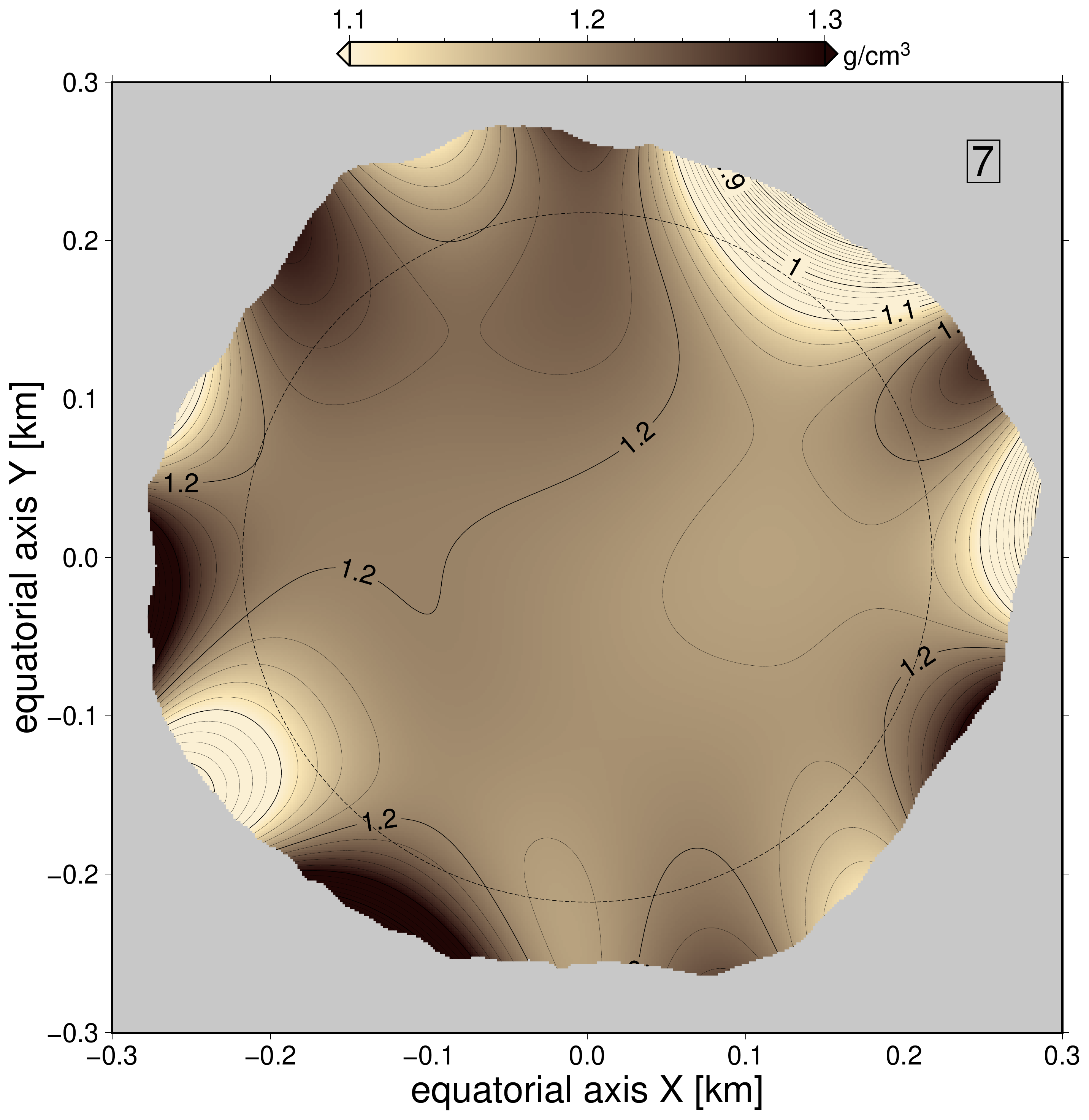}
\caption{
{\bf Interior density solutions for Bennu -- Equatorial Section.}
Same solutions as Figure~\ref{fig:slice_XZ}, but for the X--Y plane.
}
\label{fig:slice_XY}
\end{figure*}

First, we look at the GGI solutions closest to homogeneous at degrees 2 to 7
 (Figures~\ref{fig:slice_XZ}, \ref{fig:slice_YZ}, and \ref{fig:slice_XY}).
For each degree, the solution displayed is selected automatically
as having a deviation from homogeneous that is as close as possible
to the mean deviation from homogeneous for all the solutions generated at that degree.
The solutions can come very close to homogeneous at low degrees, 
and then larger deviations appear at higher degrees.
As expected, the largest deviations are concentrated near the surface,
where the spherical harmonics expansion of the gravitational field is more sensitive.
Although at low degrees (approximately 2 to 5), there is a coherence in how each additional degree 
increases the detail of the density gradient, at higher degrees (6 and above), the solution has much larger fluctuations,
and some features appear that are not present in lower-degree solutions.
This is most likely due to the fact that at higher degrees, the gravity coefficients
are determined with less than one order of magnitude accuracy \citep{Scheereseabc3350}, 
and this can affect the GGI technique, which is sensitive to the accuracy of the input gravity coefficients.
Note that the calculations use the image-based shape model v42 of Bennu (updated from v20 presented in \citet{2019NatGe..12..247B} and available via the Small Body Mapping Tool\footnote{\url{http://sbmt.jhuapl.edu/Object-Template.php?obj=77}}).

\begin{figure*}
\centering
\includegraphics*[width=0.90\textwidth]{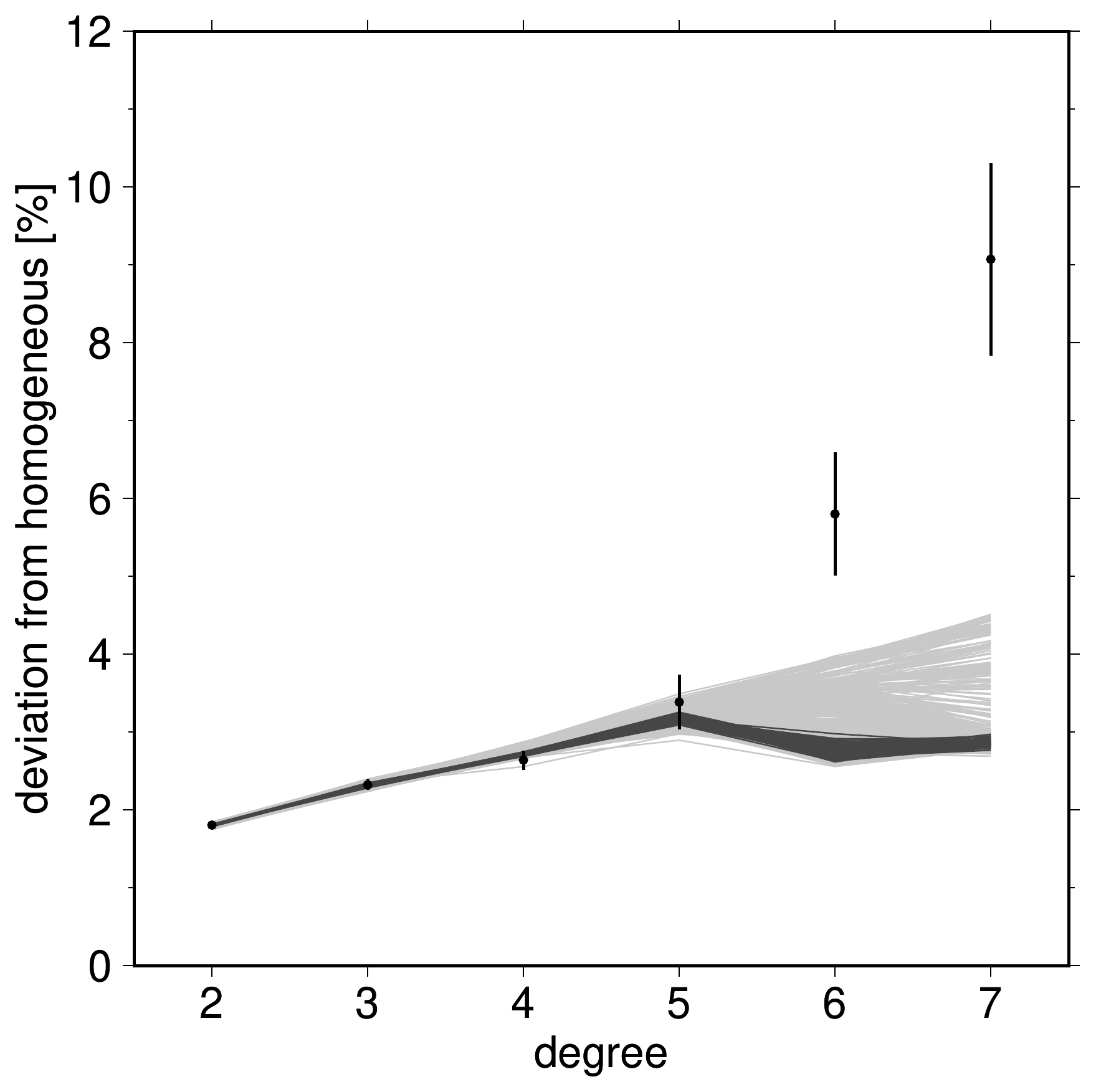}
\caption{
{\bf Bennu’s deviation from a homogeneous interior density distribution.}
The data points represent the deviation from homogeneous, measured as $\sigma_\rho/\mu_\rho$,
depending on the degree $l$ of the gravity data included in the GGI analysis.
The error bars correspond to one standard deviation of the distribution observed.
The lines are a fitting to the data using the statistical rubble modeling, including only the data for degrees 2 to 5.
The grayscale corresponds to $1\sigma$ uncertainty (dark) and $2\sigma$ uncertainty (light).
}
\label{fig:data}
\end{figure*}

A quantitative analysis of this effect is displayed in Figure~\ref{fig:data},
where for each degree we measure the deviation from homogeneous.
Up to degree 5, there is a nearly linear trend, which breaks at degrees 6 and higher.
The curves fitting degrees 2 to 5 are obtained by statistically modeling the rubble pile.
The limited number of data points that can be used for the fit requires us to keep the number of free parameters as low as possible,
leading to a model with only three:
the rubble density $\rho$, the rubble cumulative SFD index $\alpha_\text{r}$, and the void cumulative SFD index $\alpha_\text{v}$.
The minimum size is set to $s_1=10^{-3}$~m, sufficiently small to avoid any possible effects on the modeling,
and the number of bins $N=200$. For the void SFD, we use the same $s_1$ and $N$ as for the rubble population.
With 4 data points and 3 free parameters, we have $4-3=1$ degrees of freedom that we use to estimate $\chi^2$ probabilities
and confidence intervals.
In particular, we refer to $1\sigma$ uncertainty for results with $\chi^2 < 1$, corresponding to a probability of $0.683$,
and $2\sigma$ uncertainty for results with $\chi^2 < 4$, corresponding to a probability of $0.955$.

\begin{figure*}
\centering
\includegraphics*[width=0.90\textwidth]{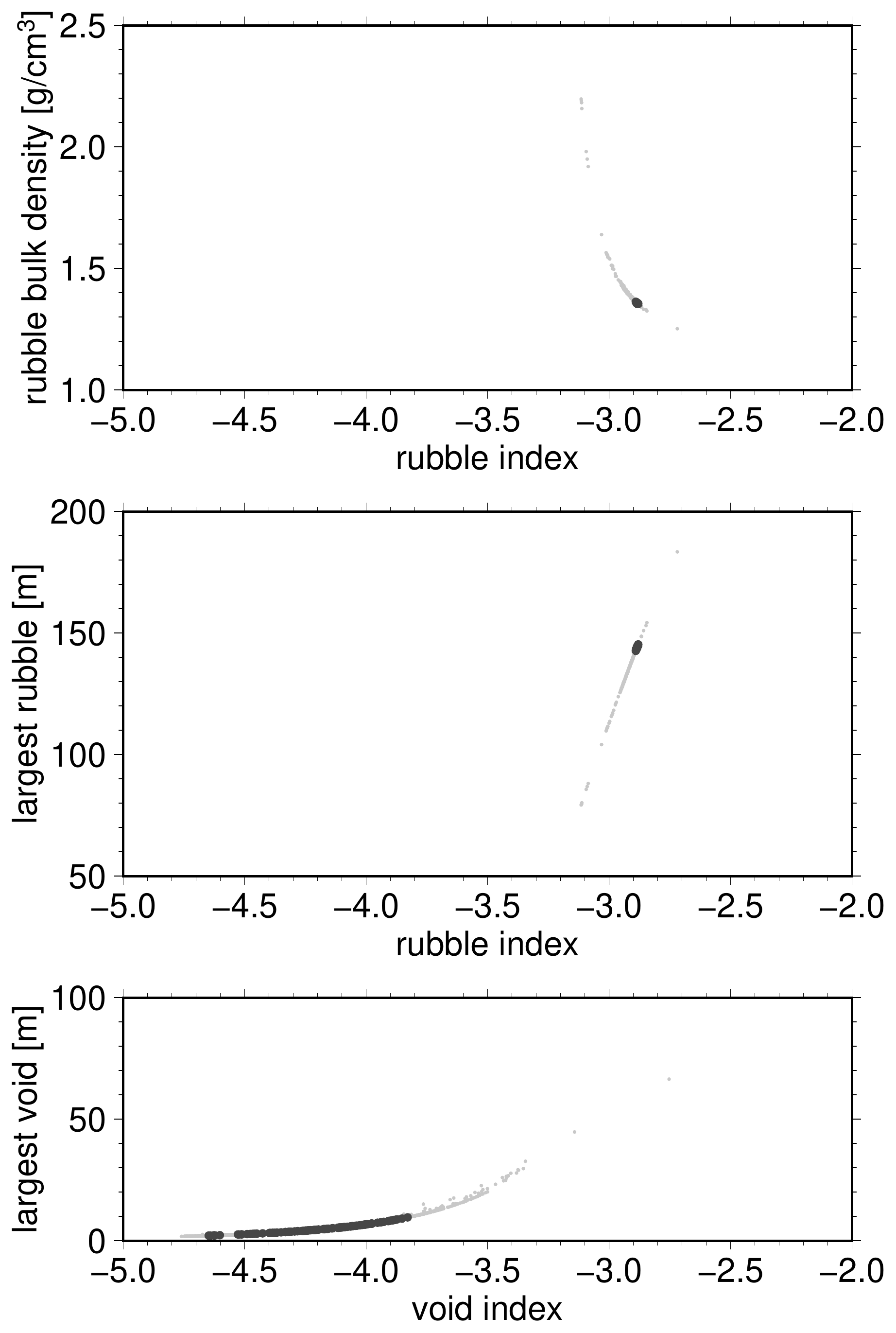}
\caption{
{\bf Properties of the rubble that constitutes Bennu.}
The diagrams show the range of rubble bulk density (top)
and largest rubble particle (middle),
as a function of the rubble cumulative SFD index.
The bottom diagram shows the size of the largest void within Bennu
as a function of the void cumulative SFD index.
The grayscale corresponds to $1\sigma$ uncertainty (dark and large points) and $2\sigma$ uncertainty (light and small points).
}
\label{fig:rubble}
\end{figure*}

In Figure~\ref{fig:rubble}, we show the range of parameters fitting the deviation from homogeneous data.
Solutions with low $\chi^2$ tend to have low rubble bulk density, between about $1.35$~g/cm$^3$ and $1.36$~g/cm$^3$ ($1\sigma$) (Figure~\ref{fig:rubble}, top panel),
compared to the bulk density of Bennu of approximately $1.19$~g/cm$^3$ \citep{2019NatGe..12..247B,ScheeresNatAstro2019}
that includes all rubble particles and voids within Bennu.

The cumulative SFD index of the rubble is between about $-2.89$ and $-2.88$ ($1\sigma$) (Figure~\ref{fig:rubble}),
in excellent agreement with the SFD index of $-2.9 \pm 0.3$ observed on the surface \citep{2019NatAs...3..341D}.
This validates the common assumption that the SFD observed on the surface can be used to model the interior.
The size of the largest rubble particle inside Bennu is approximately between 143~m and 145~m ($1\sigma$) (Figure~\ref{fig:rubble}, middle panel).
Note that we have filtered out all possible solutions where the largest rubble particle would be smaller than $50$~m,
as boulders of this size have been observed on the surface of Bennu \citep{2019NatAs...3..341D}.

The cumulative SFD index of the voids is between about $-4.6$ and $-3.8$ ($1\sigma$) (Figure~\ref{fig:rubble}, bottom panel),
much steeper than for the rubble, indicating that the number of small voids grows
at a faster pace than the number of small rubble particles.
The largest voids are about $2$~m to $10$~m ($1\sigma$).

\section{Discussion} \label{sec:discussion}

The rubble bulk density of $1.35$~g/cm$^3$ to $1.36$~g/cm$^3$ is lower than typical values for the carbonaceous chondrite types CI and CM \citep{2011M&PS...46.1842M},
which have been spectrally associated with Bennu \citep{2011Icar..216..462C}.
This may indicate a higher micro-porosity within each particle compared to what is typically observed in meteorite samples,
at least down to the smaller scale included in the modeling of the order of a millimeter.
A material with higher micro-porosity would be more likely to be destroyed during Earth atmospheric entry, thus explaining the lack of representative samples on Earth. This is consistent with the results of \citet{2020SciA....6.3699R}, who show that the dominant, low-reflectance population of boulders on Bennu \citep{2020Sci...370.3660D} has lower thermal inertia and higher porosity than expected from the meteorite record.
Our estimated macro-porosity of Bennu of approximately $12$\%
indicates relatively tight packing, and may also support the idea of fragile rubble
blocks that tend to fragment when in contact with other blocks,
helping to fill the inter-rubble voids.
The finding of relatively wide voids is a confirmation of \citet{Scheereseabc3350},
and is in agreement with earlier modeling by \citet{2001Icar..152..134B}.
Some modeling choices can also affect the results,
as is the case for the smallest rubble scale used of the order of a millimeter,
that if increased or decreased would cause a proportional correction in the estimated
size of the largest rubble particle, with a proportionality factor approximately equal to $0.003$,
while the same correction for the size of the largest void has a proportionality factor of approximately $0.1$.
Such an effect is sufficiently small for the rubble to be negligible,
while for the voids it is larger,
due at least in part to the much steeper void SFD.
The nature of the relatively wide range of void SFD index is open to speculation,
and may be difficult to interpret as voids can connect in many different ways and form complex structures.
This could simply suggest a limitation in modeling voids using an SFD,
but it could alternatively indicate the presence of voids originating in the rotational history of Bennu,
superposed to a baseline of voids following a power-law SFD.

The analysis presented here is relatively simple and avoids assumptions
about specific interior density distributions or properties of the rubble or the voids,
beyond what is strictly necessary.
This can be compared directly with \citet{Scheereseabc3350},
where we tested mass distributions with an under-dense center and equatorial bulge,
and these additional assumptions cause significantly larger deviations from a homogeneous mass distribution.

The strong agreement between the cumulative SFD index of the interior with the index determined by observing the surface 
suggests that the methods used are adequate for this class of studies
and offer the possibility to analyse other rubble-pile asteroids explored by space missions.
Although we mainly discuss $1\sigma$ uncertainty values,
at $2\sigma$, the ranges of rubble and void properties tend to be considerably wider,
see Figure~\ref{fig:rubble}.
Both the rubble SFD and the void SFD are assumed to apply globally,
a modeling simplification imposed on us principally by the limited data available,
preventing us from testing for possible effects such as size sorting or segregation mechanisms \citep{2001Icar..152..134B}.

A similar analysis of rubble-pile asteroid Ryugu is presented by \citet{Grott_2020},
who report a macro-porosity of $(16 \pm 3)$\% using the surface SFD from \citet{2019Icar..331..179M}.
Their analysis relies on detailed modeling of the rubble characteristics,
including effects from rubble size and shape distribution.
Compared to our approach, they do not use input from detailed gravity modeling,
they fix the SFD index to the value observed on the surface,
and they fix the minimum rubble size to $0.02$~m and the maximum rubble size to $140$~m.
As Ryugu has a bulk density of approximately $1.19$~g/cm$^3$, the same as Bennu \citep{2019NatGe..12..247B,ScheeresNatAstro2019},
it is interesting to note that our estimated macro-porosity of Bennu of approximately $12$\%
is also close to the macro-porosity estimated for Ryugu, even if the cumulative SFD index for Ryugu is approximately $-2.65$.

\section{Conclusions} \label{sec:conclusions}

The deviations from homogeneous of the interior mass density distribution of Bennu are consistent with the density fluctuations of a rubble pile with the same SFD index as the surface, that is, –2.9.
The bulk density of the rubble is approximately $1.35$~g/cm$^3$,
indicating a moderate macro-porosity of approximately $12$\% and implying significant micro-porosity within the constituent particles of the rubble.

\acknowledgments

We are grateful to the entire OSIRIS-REx Team for making the encounter with Bennu possible.
This material is based upon work supported by NASA under Contract NNM10AA11C issued through the New Frontiers Program.
This research is supported by the NASA OSIRIS-REx Participating Scientists Program,
grant 80NSSC18K0280 to the Planetary Science Institute.
Part of the research was carried out at the Jet Propulsion Laboratory, California Institute of Technology,
under a contract with the National Aeronautics and Space Administration (80NM0018D0004).

\clearpage

\end{document}